\documentclass[%
 reprint,superscriptaddress,
 amsmath,amssymb,prb,
]{revtex4-2}

\usepackage{graphicx}
\usepackage{dcolumn}
\usepackage{bm}
\usepackage{color}
\usepackage{siunitx}
\usepackage[version=3]{mhchem} 
\usepackage{xr}
\usepackage[normalem]{ulem}
\usepackage{soul}
\def \CCI{CeCoIn$_{5}$}

\usepackage[colorlinks=true]{hyperref}  
\hypersetup{
    unicode=false,          
    pdftoolbar=true,        
    pdfmenubar=true,        
    pdffitwindow=false,     
    pdfstartview={FitH},    
    pdftitle={FloquetAncilla},    
    pdfauthor={},     
    pdfsubject={},   
    pdfcreator={},   
    pdfproducer={}, 
    pdfkeywords={} {} {}, 
    pdfnewwindow=true,      
    colorlinks=true,       
    linkcolor=blue, 
    citecolor=blue,        
    filecolor=magenta,      
    urlcolor=blue           
} 

\begin{document}
\preprint{APS/123-QED}

\title{Superconductivity and fractionalized magnetic excitations in \CCI{}}
\thanks{This manuscript has been authored by UT-Battelle, LLC under Contract No. DE-AC05-00OR22725 with the U.S. Department of Energy.  The United States Government retains and the publisher, by accepting the article for publication, acknowledges that the United States Government retains a non-exclusive, paid-up, irrevocable, world-wide license to publish or reproduce the published form of this manuscript, or allow others to do so, for United States Government purposes.  The Department of Energy will provide public access to these results of federally sponsored research in accordance with the DOE Public Access Plan (http://energy.gov/downloads/doe-public-access-plan).}

\author{Pyeongjae Park}
\email{parkp@ornl.gov}
\affiliation{Materials Science \& Technology Division, Oak Ridge National Laboratory, Oak Ridge, TN 37831, USA}

\author{Shang-Shun Zhang}
\affiliation{Department of Physics and Astronomy, The University of Tennessee, Knoxville, TN, 37996, USA}

\author{Pietro M. Bonetti}
\affiliation{Department of Physics, Harvard University, Cambridge, MA 02138, USA}
\affiliation{Max Planck Institute for Solid State Research, Heisenbergstraße 1, D-70569 Stuttgart, Germany}

\author{Andrey A. Podlesnyak}
\affiliation{Neutron Scattering Division, Oak Ridge National Laboratory, Oak Ridge, Tennessee 37831, USA}

\author{Daniel M. Pajerowski}
\affiliation{Neutron Scattering Division, Oak Ridge National Laboratory, Oak Ridge, Tennessee 37831, USA}

\author{Matthew B. Stone}
\affiliation{Neutron Scattering Division, Oak Ridge National Laboratory, Oak Ridge, Tennessee 37831, USA}

\author{C. Petrovic}
\affiliation{Department of Physics, Brookhaven National Laboratory, Upton, New York 11973, USA}

\author{C. Stock}
\affiliation{School of Physics and Astronomy, University of Edinburgh, Edinburgh EH9 3JZ, United Kingdom}

\author{Subir Sachdev}
\affiliation{Department of Physics, Harvard University, Cambridge, MA 02138, USA}
\affiliation{Center for Computational Quantum Physics, Flatiron Institute, 162 5th Avenue, New York, NY 10010, USA}

\author{Cristian D. Batista}
\email{cbatist2@utk.edu}
\affiliation{Department of Physics and Astronomy, The University of Tennessee, Knoxville, TN, 37996, USA}
\affiliation{Quantum Condensed Matter Division and Shull-Wollan Center, Oak Ridge National Laboratory, Oak Ridge, TN, 37831, USA}

\author{Andrew D. Christianson}
\email{christiansad@ornl.gov}
\affiliation{Materials Science \& Technology Division, Oak Ridge National Laboratory, Oak Ridge, TN 37831, USA}

\begin{abstract}

Recent experiments on CeCoIn$_5$---a prototypical $d$-wave superconductor---indicate that its normal state lies near an unconventional quantum critical point (QCP). One intriguing hypothesis is that quantum-critical fluctuations promote fractionalization of localized $4f$ moments into fermionic spinons. This fractionalized Fermi liquid (FL$^{*}$) scenario provides a comprehensive framework for the unconventional QCP and superconductivity, and can reconcile a “missing” Fermi-surface volume relative to the Luttinger count in the normal state of CeCoIn$_5$. To test this possibility, we performed inelastic neutron scattering (INS) measurements on CeCoIn$_{5}$ across the superconducting transition and corresponding theoretical analysis. Our high-precision spectra reveal detailed momentum and temperature dependence of the spin resonance and a structured spin excitation continuum persisting even in the normal state, placing stringent constraints on the physical picture of pairing in a $d$-wave superconductor. We show that a Kondo-lattice framework incorporating proximity to FL$^*$ physics and $d$-wave pairing reproduces key features of the data. The model suggests that both the quasi-localized nature of the $f$-moments above $T_{\mathrm{c}}$ and the resonance below $T_{\mathrm{c}}$ arise from common underlying gauge dynamics, implying a unifying organizing principle linking spin fractionalization and unconventional superconductivity in strongly correlated metals. 

\end{abstract}

\maketitle
Strong electronic correlations, which can lead to phenomena beyond the Landau Fermi-liquid paradigm, are widely believed to underlie unconventional superconductivity. The heavy-fermion metal CeCoIn$_5$ is a prototypical $d$-wave superconductor, which lies in close proximity to an unconventional quantum critical point (QCP)~\cite{Petrovic2001,Settai2001,Kohori2001,Sparn2002,Sidorov2002,Bianchi2003,Paglione2003,Christianson2004,Nakajima2007,stock2008,ramos2010,Zhou2013,Tokiwa2013,Allan2013,VanDyke2014,Spehling2009}. Previous experiments have revealed pronounced non-Fermi liquid behavior at zero field and a field-driven evolution into a conventional (non-magnetic) Fermi liquid as well as a magnetically ordered phase~\cite{Paglione2003,Bianchi2003,Tokiwa2013, kenzelmann2008,kenzelmann2010}, placing CeCoIn$_{5}$ near a magnetic QCP separating antiferromagnetically ordered and heavy Fermi-liquid regimes. Notably, this magnetic criticality appears incompatible with the conventional Hertz-Millis-Moriya description based on spin-density-wave instability, pointing instead to alternative explanations such as a Kondo-breakdown scenario~\cite{Gegenwart2008,si2010,stockert2011}.

An intriguing hypothesis is that the unconventional QCP of CeCoIn$_{5}$ is characterized by the fractionalization of the localized $4f$ magnetic moments into fermionic spinons. This fractionalized Fermi liquid (FL$^*$) scenario offers a concrete framework for understanding Kondo-breakdown quantum criticality~\cite{senthil2003, senthil2004, paramekanti2004, Andrei1989, Coleman1989, christos2023model, christos2024emergence, bonetti2025critical}. Notably, this can explain a few puzzling observations in CeCoIn$_{5}$. While extensive experiments have demonstrated signatures of Kondo hybridization at low temperatures~\cite{shishido2002, koitzsch2008, koitzsch2013, chen2017}, observations of Maksimovic {\it et al.\/} and others \cite{maksimovic2022, koitzsch2009, chen2017} indicate a partially ``missing" Fermi-surface volume in the normal state relative to the Luttinger count, without any symmetry-breaking long-range order~. This apparent violation can be reconciled in FL$^*$ by assigning the missing volume to the fractionalized, charge-neutral spin sector \cite{senthil2004,paramekanti2004,Bonderson16,SenthilElse21,Seiberg23}, compatible with a more generalized form of the Luttinger constraint formulated by Oshikawa~\cite{oshikawa2000}. Importantly, this parton description does not require the system to lie exactly at the QCP: even if the ultimate low-temperature state is confined, proximity to a deconfined QCP may still yield the deconfinement length significantly exceeding the lattice spacing, making the parton framework a natural starting point for finite-temperature dynamics. This is analogous to magnetically ordered systems proximate to quantum spin liquids where collective modes are more accurately described as two-spinon bound states rather than conventional magnons~\cite{Ghioldi2018,Ghioldi2022}.

A central outstanding question then is whether CeCoIn$_5$ shows hallmarks of spin fractionalization in its normal state, and whether these fractionalized excitations confine in the superconductor to yield a state smoothly connected to a Bardeen-Cooper-Schrieffer state. Inelastic neutron scattering (INS) is uniquely suited to address this issue. Beyond establishing proximity to deconfined spinons in quantum spin-liquid candidates~\cite{lake2005, banerjee2017, scheie2024}, INS has historically revealed key collective excitations in unconventional superconductors, most notably the sharp, gapped ``spin-resonance’’ modes that emerge below $T_{\mathrm{c}}$ at antiferromagnetic wave vectors in cuprates, iron pnictides, and heavy-fermion materials~\cite{sidis2004, lumsden2010, fujita2011, tranquada2014, dai2015, stockert2023}. In CeCoIn$_5$, seminal INS studies identified such a resonance at $\mathbf{Q}_{\delta} = (1/2 - \delta,\, 1/2 - \delta,\, 1/2)$ with $\delta \approx 0.05$, and traced its evolution into the field-induced, incommensurate ``Q-phase’’ spin-density wave governed by the same $\mathbf{Q}_{\delta}$~\cite{stock2008, kenzelmann2008, kenzelmann2010, stock2012, raymond2012, raymond2015, song2016, song2020}. However, these measurements were largely confined to a narrow region around $\mathbf{Q}_{\delta}$ and to temperatures below $T_{\mathrm{c}}$. Mapping the complete momentum- and energy-dependent evolution of this mode across the superconducting transition could offer essential clues to its microscopic origin and, critically, to whether it reflects the emergence of fractionalized spin excitations potentially underlying a nearby magnetic QCP.

\begin{figure*}[ht!]
\includegraphics[width=1\textwidth]{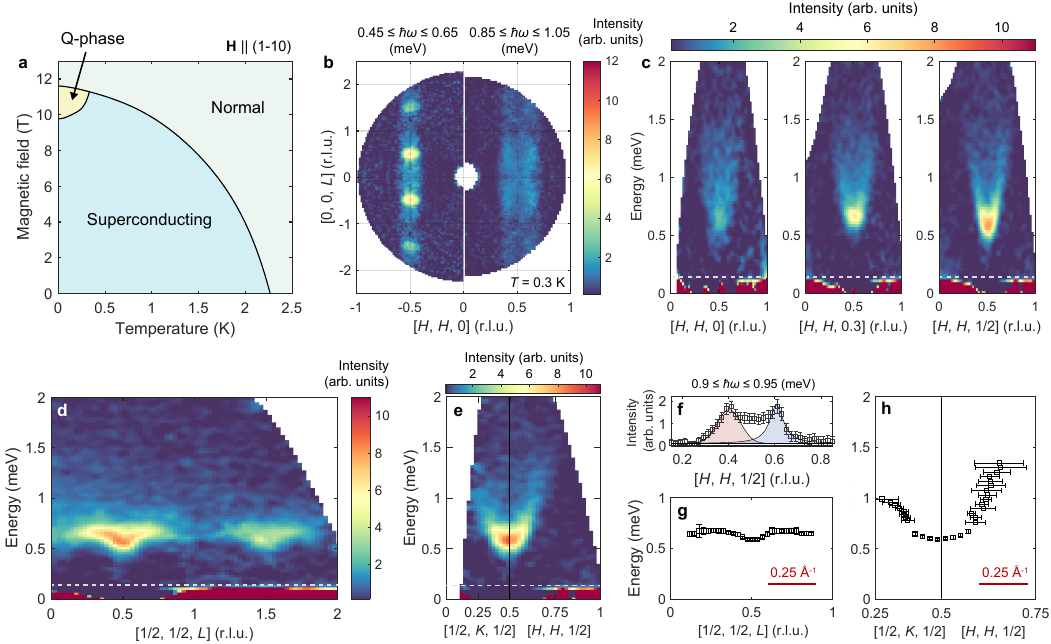} 
\caption{\label{INS_SC} \textbf{Magnetic excitation spectra in the superconducting phase of \CCI{} (\textit{T} = 0.3\,K).} \textbf{a}, Field–temperature phase diagram of \CCI{}~\cite{kenzelmann2008, kenzelmann2010}. The ``Q-phase" corresponds to a spin-density wave order of Ce$^{3+}$ moments with modulation vector $\mathbf{Q}_{\delta} = (0.45, 0.45, 1/2)$ in reciprocal lattice units. \textbf{b}, Two constant-energy slices of the inelastic neutron scattering (INS) spectrum at $T = 0.3$\,K, integrated over two different energy ranges. \textbf{c}--\textbf{e}, Energy--momentum slices along high-symmetry directions, revealing a well-defined magnetic excitation branch across broad momentum space as well as continuum scattering. Dashed grey lines mark regions where the spectrum is affected by unreliable background subtraction of quasi-elastic scattering signals (see Fig.~S3). \textbf{f} Momentum-dependent intensity over the energy range $0.9 \leq \hbar\omega \leq 0.95$\,meV at $T = 0.3$\,K. The red and blue shaded regions indicate nominal peaks obtained by fitting each side of the dispersive mode with a pseudo-Voigt function. \textbf{g}--\textbf{h} Extracted dispersion relations of the observed excitation branch from panels \textbf{d} and \textbf{e}, obtained by fitting the peak positions of constant-energy/momentum cuts through the volumetric intensity data.} 
\end{figure*}

Here, using a modern time-of-flight spectrometer with broad momentum coverage and enhanced statistics, we report comprehensive INS measurements of CeCoIn$_5$. These measurements provide a significantly improved view of the magnetic excitation spectrum across the superconducting transition and uncover previously inaccessible features: namely, quasi two-dimensional spin dynamics, a scattering continuum co-existing with coherent gapped excitations, and strong antiferromagnetic spin correlations that persist above $T_{\mathrm{c}}$. These points establish a stringent experimental benchmark for theories of correlated superconductivity.  We utilize the spectroscopic data to test the theoretical framework incorporating both the proximity to a FL$^*$ phase and $d$-wave pairing, which is based on a fermionic parton decomposition of the $4f$-moments consisting of an emergent SU(2) gauge field coupled to $N_f=2$ flavors of Dirac fermions. The model captures the overall energy, momentum, and temperature dependence of the experimental spectrum and thus supports a link between spin fractionalization and unconventional superconductivity in strongly correlated metals.

\section*{Experiment}
Our dataset, which spans both the superconducting and normal phases, reveals a transformation of the magnetic excitation spectrum across $T_{\mathrm{c}}$. Below $T_{\mathrm{c}}$ (Fig. 1), the most prominent feature is a well-defined, dispersive magnetic excitation branch with a finite spin gap, extending continuously along both in-plane and out-of-plane momentum directions. While this excitation indeed connects to the previously reported spin resonance mode~\cite{stock2008}, our full mapping along the high-symmetry directions—$[H, H, 0]$, $[H, 0, 0]$, and $[0, 0, L]$—allows us to resolve its detailed dispersion in each direction (Figs.~1g--h).

\begin{figure*}[ht!]
\includegraphics[width=1\textwidth]{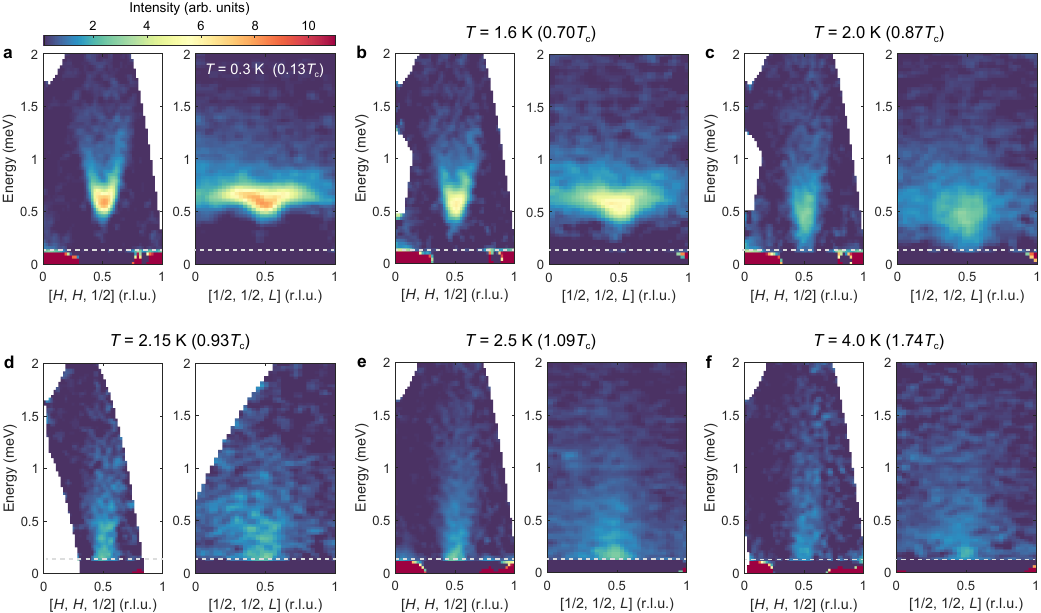} 
\caption{\label{INS_Tdep} \textbf{Temperature dependence of the magnetic excitations in \CCI{}}. Each panel shows energy-momentum slices along [1/2, 1/2, $L$] and [$H$, $H$, 1/2] measured at \textbf{a} $T= 0.3$\,K, \textbf{b} 1.6\,K, \textbf{c} 2\,K, \textbf{d} 2.15\,K, \textbf{e} 2.5\,K, and \textbf{f} 4\,K. They demonstrate the evolution of the gapped magnetic excitations into a gapless, structured continuum scattering signal above $T_{\mathrm{c}}$. Dashed grey lines mark regions where the spectrum is affected by unreliable background subtraction due to strong quasi-elastic background signals (see Fig.~S3).} 
\end{figure*}

Notably, the dispersion along [0, 0, $L$] is nearly flat compared to the in-plane directions, highlighting the quasi-two-dimensional nature of the spin dynamics. This contrasts with earlier reports of nearly isotropic dispersion~\cite{song2016} and provides motivation for modeling the spin dynamics using a two-dimensional square-lattice framework. In a similar context, the gapped spin excitation previously thought to be present only around $L = 1/2$ is in fact not strictly confined to the $L=1/2$ plane. Instead, it extends across nearly the entire range of $L$ values (Fig.~1c--d). In addition, the intensity of this mode decreases with increasing $L$ (e.g., [1/2, 1/2, 1/2] and [1/2, 1/2, 3/2] in Fig.~\ref{INS_SC}d), consistent with prior reports that the spin fluctuations are predominantly $c$-axis polarized, as the momentum transfer becomes more parallel to the $c$-axis at higher $L$~\cite{stock2008, raymond2015}.

More importantly, beyond the coherent dispersive branch, our measurements reveal a broad underlying magnetic continuum. As best illustrated in Fig.~\ref{INS_SC}f, a substantial portion of the spectral weight resides within the dispersion valley, well beyond what can be captured by the model of a single, coherent excitation branch with V-shaped dispersion. This spectral weight exhibits a broad distribution that extends well beyond the instrumental momentum and energy resolutions (see Supplementary Note II), supporting its continuum-like character. See also the constant-energy slices above 0.85\,meV in Fig.~\ref{INS_SC}b and Fig.~S5. The coexistence of a sharp, dispersive mode and a broad continuum suggests correlated spin dynamics beyond a simple single-particle picture, likely reflecting two-particle dynamics of underlying $S = 1/2$ degrees of freedom. Notably, this continuum character becomes increasingly pronounced as the [0, 0, $L$] component deviates from $L = 1/2$ (Figs.~1c--d).

The temperature dependence of these excitations provides further insight into the close link between superconductivity and spin correlations in CeCoIn$_{5}$. As the temperature rises through $T_{\mathrm{c}}$, the gapped, dispersive mode collapses into a broad, gapless continuum extending down to the lowest accessible energy ($\sim0.15$\,meV; see Fig.~\ref{INS_Tdep}). This transformation indicates that the entire coherent branch, not just the resonance near $\mathbf{Q}_{\delta}$, is closely tied to the onset of superconductivity. Remarkably, in the normal state, the continuum remains well structured, with spectral weight concentrated near $\mathbf{Q}_{\delta}$, indicating persistent antiferromagnetic correlations. This challenges the prevailing view that the magnetic spectrum of CeCoIn$_5$ above $T_{c}$ is featureless, which led to its routine subtraction as “background” in prior neutron studies~\cite{stock2008, song2020}. This inadvertently obscured an intrinsic continuum component that coexists with the coherent mode below $T_{\mathrm{c}}$ (see Fig.~\ref{INS_SC}f).

In addition, the spectral weight near the resonance rapidly diminishes approaching $T_{\mathrm{c}}$, and becomes significantly weaker at higher temperatures. Fig.~S6 shows detailed line cuts at $\mathbf{q}=(1/2,1/2,1/2)$, illustrating this evolution together with Fig.~\ref{INS_Tdep}. Although our analysis may miss some spectral weight below $\sim0.15$\,meV in the normal phase due to background contamination, a quantitative assessment of the total intensity at $\mathbf{q}=(1/2,1/2,1/2)$ likely suggests a strong net enhancement of the structure factor below $T_{\mathrm{c}}$, rather than a redistribution of spectral weight from zero to finite energy at constant total intensity.

Overall, these results suggest that the magnetic excitation spectrum of CeCoIn$_5$ is far richer than a single superconductivity-induced ``spin resonance.’’ It reveals a continuous evolution from a scattering continuum---rooted in strong antiferromagnetic correlations of the normal state---into a coherent, gapped mode that develops alongside superconductivity. The intertwined nature of these excitations highlights the need for a theoretical framework that can capture both coherent and incoherent magnetic responses across energy--momentum space, by treating antiferromagnetic correlations and superconductivity on equal footing.

\section*{Theory}

\begin{figure*}[ht!]
\includegraphics[width=0.9\textwidth]{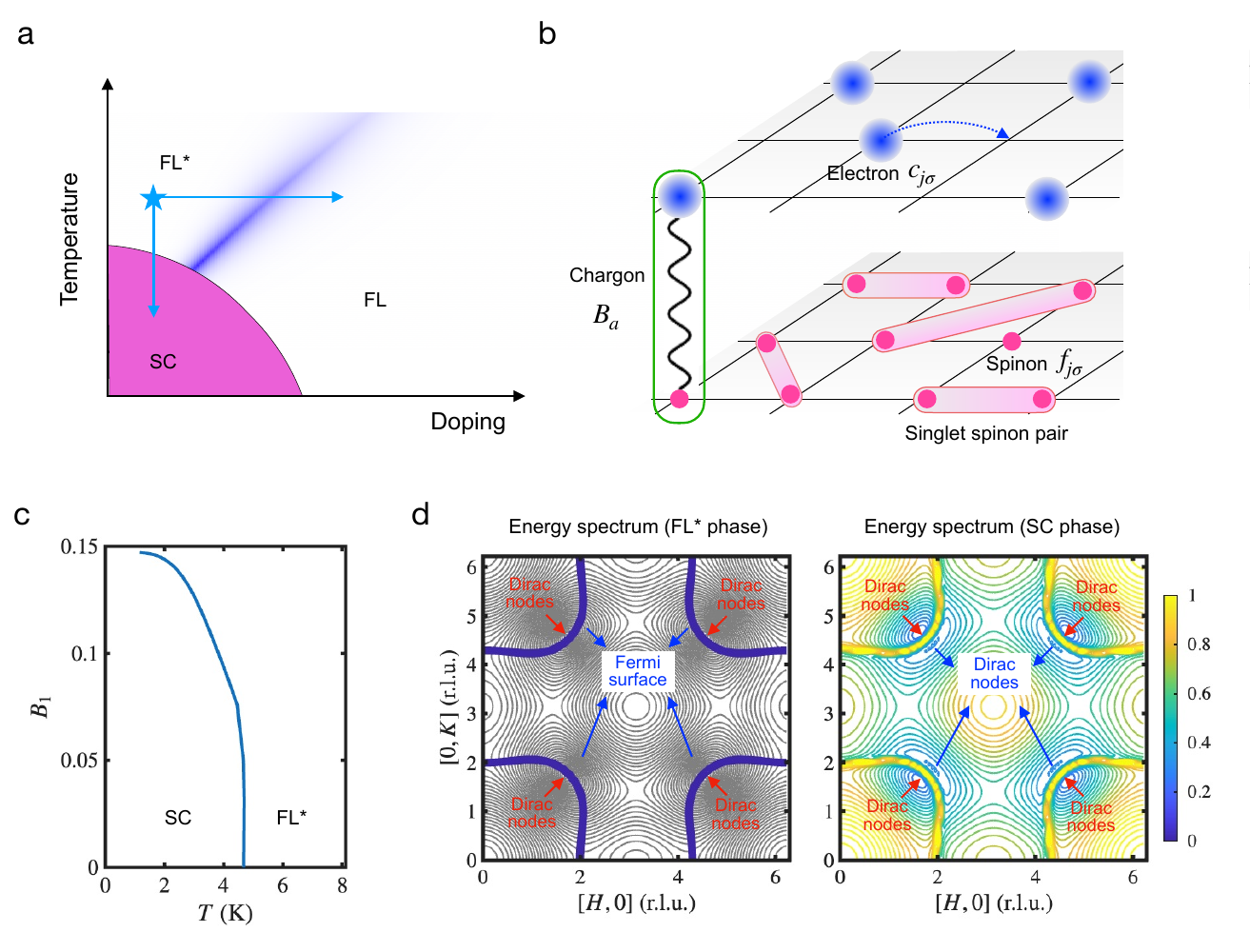} 
\caption{\label{model} \textbf{Minimal theoretical model for CeCoIn$_5$.} 
\textbf{a}, Schematic theoretical phase diagram of CeCoIn$_5$ in the temperature–doping plane. Upon doping (horizontal arrow), the system undergoes a delocalization transition, as reported in Ref.~\cite{maksimovic2022}. Upon cooling (vertical arrow), it enters a superconducting state through the condensation of chargons. 
\textbf{b}, The minimal model capturing the proposed scenario consists of conduction electrons and local $f$-moments, coupled via a Kondo-like interaction. A central assumption is that the $f$-moments fractionalize into fermionic spinons, giving rise to a fractionalized FL* phase above $T_c$.
A bosonic chargon field emerges as a bound state of an electron and a spinon, and its condensation drives the onset of superconductivity.
The corresponding chargon condensate order parameter, $B_1$, is shown in \textbf{c} as a function of temperature.
\textbf{d,} Contour plot of the low-energy spectrum in the FL$^*$ and superconducting phases. The color scale of the contour plot represents energy of the single-particle excitations. The gray contours indicates that individual spinon in the FL* phase is invisible to physical probes.
} 
\end{figure*}

Parton-based approaches provide an intuitive framework for describing the continuum of magnetic excitations that emerges in deconfined quantum phases~\cite{christos2023model,christos2024emergence,bonetti2025critical,punk2014topological} or in their proximity~\cite{Ghioldi2018,Ghioldi2022}, where the deconfinement length scale remains finite but much larger than the lattice spacing. The finite-temperature FL$^*$ state---a compelling framework for the normal phase of CeCoIn$_{5}$---represents such a regime, in which the evolution toward a fully deconfined zero-temperature phase is interrupted by the onset of $d$-wave superconductivity that confines the emergent gauge fields via the Higgs mechanism (see Fig.~\ref{model}a and b) \footnote{Although the deconfinement associated with the FL$^*$ phase exists strictly at zero temperature, we continue to refer to the finite-temperature normal state as FL$^*$ for brevity.}. This raises a natural question: can the continuous transition between the FL$^*$ and superconducting states, as captured by a simple mean-field parton theory, account for the INS spectra of CeCoIn$_{5}$ below and above $T_{\mathrm{c}}$? To address this question, we describe CeCoIn$_5$ using an effective Kondo–Heisenberg lattice model:
\begin{equation}
    {\cal H} = {\cal H}_c + {\cal H}_f + {\cal H}_K,
\end{equation}
with
\begin{eqnarray}
{\cal H}_c &=& \sum_{ij} \left( t_{ij} c_{i\sigma}^\dagger c_{j\sigma} + \text{H.c.} \right) - \mu \sum_{i,\sigma} c_{i\sigma}^\dagger c_{i\sigma},
\nonumber \\
{\cal H}_f &=& J_H \sum_{\langle ij \rangle} {\bf S}_i \cdot {\bf S}_j, \quad {\cal H}_K = \frac{1}{2} J_K \sum_i {\bf S}_i \cdot c_i^\dagger \boldsymbol{\sigma} c_i.
\end{eqnarray}
${\cal H}_c$ denotes the tight-binding Hamiltonian for the conduction electrons, including hopping amplitudes up to third-nearest neighbors, whose values are determined by fitting quasiparticle interference (QPI) data~\cite{allan2013imaging,van2014direct}. ${\cal H}_f$ describes the effective interactions among the localized $f$-moments, while ${\cal H}_K$ captures the Kondo exchange coupling between the $f$-electron spins and the conduction electrons.

We consider an FL$^*$ framework in which the Ce$^{3+}$ 4$f$ moments fractionalize into fermionic spinons, while the conduction electrons retain a conventional Fermi-liquid character. The spinon degrees of freedom are introduced by expressing the spin-exchange interaction in bilinear form, ${\bf S}_i = \tfrac{1}{2} f_i^\dagger {\boldsymbol \sigma} f_i$, with $f_i = (f_{i,\uparrow}, f_{i,\downarrow})^T$. The resulting spin-liquid state is described at the mean-field level by the quadratic Hamiltonian
\begin{equation}
    {\cal H}_{f}^{\rm m.f.} = \frac{J_H}{4}\sum_{\langle jj' \rangle }\left(F_j^\dagger U_{jj'}^{\rm m.f.} F_{j'} + {\rm H.c.}\right),
\end{equation}
where $\langle jj' \rangle $ refers to nearest neighbor bonds, $F_j = (f_{j,\uparrow}, f_{j,\downarrow}^\dagger)^T$ is the Nambu spinor, and 
\begin{equation}
    U_{jj'}^{\rm m.f.} =
    \begin{pmatrix}
        t & e_{jj'} \Delta \\
        e_{jj'} \Delta & -t
    \end{pmatrix}
\end{equation}
is the mean-field ansatz in the standard $d$-wave gauge, with $e_{jj'} = +1\,(-1)$ for bonds along the $x$\,($y$) direction. This ansatz describes a spin-liquid state with an emergent ${\mathrm U}(1)$ gauge symmetry, which enlarges to SU(2) at the special point $t = \Delta$ for which there is support in numerical studies of square lattice antiferromagnets \cite{Wang17,Fuzzy24,Meng24,Gu24,Chester24,Sandvik24}. In more general settings, where pairing terms on further-neighbor bonds are included, the gauge symmetry is reduced to ${\mathrm Z}_2$, corresponding to the projective symmetry group ${\mathrm Z}_2{\rm Azz13}$ \cite{WenPSG,ShackletonZ2}. We note that the specific choice of gauge structure does not qualitatively affect the dynamical spin correlations discussed below. For simplicity, we therefore set $t=\Delta$. In this class of spin liquids, the spinon spectrum is characterized by the presence of four Dirac nodes (see Fig.~\ref{model}d). 
As shown below, this mean-field description, when supplemented by quantum fluctuations, provides a consistent account of the neutron-scattering data.

The Kondo interaction acts as a pairing mechanism between spinons and conduction electrons~\cite{Andrei1989,Coleman1989}. As shown below, these two elementary excitations of the FL$^*$ phase bind to form a composite boson---referred to as a \emph{chargon}---that carries both the physical ${\mathrm U}(1)$ charge and the emergent gauge charge of the spinons (Fig.~\ref{model}b). Formally, the Kondo coupling can be rewritten as
\begin{equation}
    {\cal H}_K = \frac{J_K}{4} \sum_j \left(F_j^\dagger B_j C_j + {\rm H.c.}\right),
\end{equation}
where $C_j = (c_{j,\uparrow},\, c_{j,\downarrow}^\dagger)^T$ is the Nambu spinor of conduction electrons, and
\begin{equation}
    B_j =
    \begin{pmatrix}
        B_{1,j} & B_{2,j} \\
        B_{2,j}^* & -B_{1,j}^*
    \end{pmatrix}
\end{equation}
is the chargon field. Above $T_c$, chargons remain uncondensed and the system resides in the FL$^*$ phase; but the chargon fluctuations are expected to be overdamped and classical, as in a recent analysis for the cuprates \cite{Pandey25}. Upon cooling, chargon condensation breaks the emergent gauge symmetry, producing conventional symmetry-breaking orders and hybridized conduction--spinon bands (Fig.~\ref{model}c). A superconducting state emerges when $B_{1,j}=B_1\neq 0$ and $B_{2,j}=0$, yielding a pairing gap proportional to $|B_1|$. The resulting superconducting order inherits its $d$-wave structure from the underlying spin-liquid correlations of the FL$^*$ state, causing the conduction-electron Fermi surface to collapse into four Dirac nodes in the Bogoliubov excitation spectrum, as illustrated in Fig.~\ref{model}d.

\begin{figure*}[ht!]
\includegraphics[width=1\textwidth]{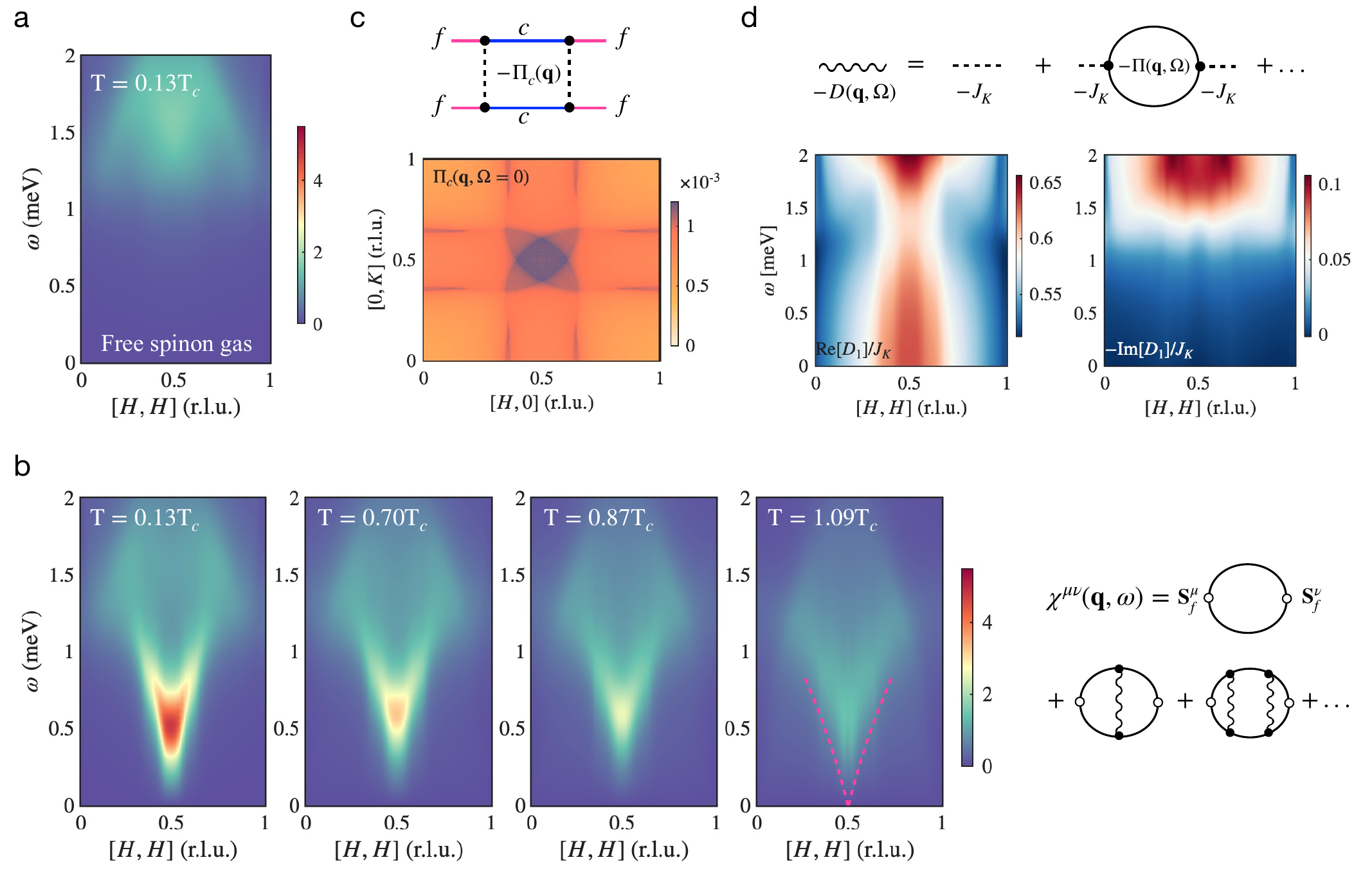} 
\caption{\label{dssf} \textbf{Calculated dynamical spin structure factor (DSSF).} \textbf{a,} DSSF between local $f$-moments in the mean-field approximation, evaluated at a temperature below $T_c$. The spectrum is characterized by a broad continuum arising from particle-hole excitations. \textbf{b,} Interaction-renormalized DSSF between local $f$-moments, obtained by summing ladder-type Feynman diagrams. A bound state of spinon pairs (or electron pairs below $T_c$) emerges below the continuum, manifesting as bright, dispersive modes. As temperature increases, this bound state gets closer to the continuum and gradually fades out. \textbf{c,} Static magnetic susceptibility $\chi(\bm{q}, 0)$ of non-interacting conduction electrons, which exhibits a peak near $(\pi,\pi)$—coinciding with the minimum of the bound-state dispersion. \textbf{d,} Renormalized propagator of the chargon field captured by the RPA series of diagrams. Here we only show the results for $B_{j,1}$ field, whose amplitude is reduced by a factor about $0.6$ relative to the bare value $J_K$; similar result holds for the $B_{j,2}$ field. 
} 
\end{figure*}

To test if this framework captures the key features in our INS data, we calculated the dynamical spin structure factor (DSSF) in the superconducting state. Given that the spectral weight is dominated by the $f$-electron moments (see Methods for details), we neglect the contribution of conduction electrons to the DSSF in our calculations. First, the spectrum obtained from a mean-field calculation (Fig.~\ref{dssf}a) exhibits a broad particle--hole continuum of the hybridized fermionic bands centered near $1.5$~meV, lacking the sharp dispersive mode observed experimentally. Intuitively, reproducing such a mode requires a quasiparticle excitation carrying spin quantum number $S_z=\pm1,0$, as dictated by magnetic dipole selection rules in neutron scattering. This type of excitation can arise from a particle--hole (i.e., two-spinon) bound state or resonance. Although this mechanism is reminiscent of the conventional spin-excitonic picture proposed to explain the spin resonance~\cite{eremin2008feedback,michal2011field,song2020}, a crucial distinction here is that both the particle and hole excitations retain substantial spinon character, reflecting the nonlocal structure of the parent FL$^*$ state. Notably, the conventional spin-exciton framework has failed to reproduce the observed mode dispersion near $\mathbf{q} = [1/2, 1/2, 1/2]$ below $T_{\mathrm{c}}$~\cite{eremin2008feedback, song2020}, and is incompatible with the non-Luttinger Fermi volume observed above $T_{\mathrm{c}}$~\cite{maksimovic2022, chen2017}.

To quantitatively assess this bound-state picture, we computed the DSSF using the ladder-diagram summation shown in Fig.~\ref{dssf}b. The bound state originates from the bubble diagrams in the ladder series, which generate Ruderman--Kittel--Kasuya--Yosida (RKKY) interaction between the local $f$ moments mediated by the conduction electrons (Fig.~\ref{dssf}c). Within second-order perturbation theory in $J_K$, this interaction takes the form $J_{\mathrm{RKKY}}({\bf q}) = J_K^2 \Pi_c({\bf q}, \Omega=0)$. As CeCoIn$_5$ lies in close proximity to a magnetically ordered phase with ordering vector $\mathbf{Q}_{\delta}$, any realistic model should exhibit $J_{\mathrm{RKKY}}({\bf q})$ that peaks near $\mathbf{Q}_{\delta}$. By tuning the conduction-electron filling $n_c$ of the single-band model, we find that $J_{\mathrm{RKKY}}({\bf q})$ is indeed maximized around ${\bf q}={\bf Q_{\delta}}$ when $n_{c} \simeq 1.2$, i.e., near half filling. The corresponding Fermi surface is close to those observed experimentally~\cite{maksimovic2022, allan2013imaging, van2014direct}. We therefore adopt $n_c=1.2$ in our calculations, and the corresponding $\Pi_c({\bf q},\Omega=0)$ is plotted in Fig.~\ref{dssf}c.

This peak in $J_{\mathrm{RKKY}}({\bf q})$ naturally leads to the spectrum with a sharp mode shown in Fig.~4b ($T=0.13T_{\mathrm{c}}$), which reproduce the key experimental features: a parabolic quasiparticle mode coexisting with a broad continuum, and a pronounced spectral weight accumulation near its dispersion minimum, i.e., the spin-resonance. In this context, the spin-resonance can be interpreted as a precursor to the field-induced magnetic instability (``$Q$-phase") that arises when this bosonic mode condenses, a perspective consistent with previous neutron scattering studies under applied magnetic fields~\cite{kenzelmann2008,kenzelmann2010,stock2012,raymond2015}.

The binding energy of the bound state is governed by the competition between the Kondo interaction $J_K$ and the Fermi energy of conduction electrons. To reproduce the experimental energy scale, it is necessary to account for the screening of the Kondo interaction, described by a momentum- and frequency-dependent RPA propagator of the $B$ fields shown in Fig.~\ref{dssf}d. Averaging this propagator over momentum and a finite frequency window ($\pm 2$meV) reduces its effective strength to approximately 60\,\% of the bare value set by $J_K$. For computational simplicity, we approximate this screened propagator as momentum- and frequency-independent when evaluating the DSSF, which yields the spectrum shown in Fig.~\ref{dssf}b ($T=0.13T_{\mathrm{c}}$). The result is in quantitative agreement with our INS measurements exhibiting the spin-resonance energy $E_r\approx0.55$\,meV. Moreover, the ratio between $E_r$ and the pair-breaking energy (= twice the superconducting gap) is close to $0.55$ in our model calculation, close to the empirical value of $\sim0.64$ that is ubiquitously observed in most unconventional superconductors exhibiting spin-resonance, including CeCoIn$_{5}$~\cite{yu2009}.

Finally, the simulated spin-resonance signal progressively weakens as the temperature increases, while the overall structure of the continuum remains largely unchanged (Fig.~\ref{dssf}b). In our framework, this reflects a reduction in the binding energy of the particle-hole bound state and a corresponding growth in its spatial extent. As $T$ approaches $T_\mathrm{c}$, the lower edge of the particle-hole continuum collapses to zero energy due to the superconducting gap closure and the presence of gapless Dirac spinons. The resonance mode is therefore embedded within a gapless continuum and loses its coherence, as shown in the rightmost panel of Fig.~\ref{dssf}b. The resulting conical continuum in the same figure (red dashed line) directly mirrors the Dirac spectrum of the underlying spin liquid above $T_c$. Notably, spinon-spinon interactions enhance low-energy spectral weight, similar to the INS data (Fig.~\ref{INS_Tdep}e). In contrast, non-interacting Dirac spinons yield vanishing spectral weight at low-energies due to the marginal density of states on the Dirac points, in qualitative disagreement with experimental observations.

\section*{Discussion and Outlook}

While our simplified model captures key features of the INS spectra---especially the spin resonance below $T_{\mathrm{c}}$ and antiferromagnetic correlations present above it---it does not fully reproduce several finer aspects, such as a steeper slope of the cone delineating the scattering continuum just above $T_c$ in the experiment (compared to the red dashed line in Fig.~\ref{dssf}b). We attribute these discrepancies to the minimal mean-field treatment of the quantum spin liquid, which neglects longer-range effective interactions between the $f$ moments that are expected to enhance the spinon velocity of the Dirac spin liquid. Incorporating these interactions substantially increases the excitation velocity while retaining a fixed bandwidth as recently demonstrated in Refs.~\cite{simeth2023microscopic, ghioldi2024derivation}, which notably improved quantitative agreement with the INS spectra of the magnetically ordered parent compound CeIn$_3$. Such a characteristic is reminiscent of the vertical pillar-like low-energy continuum structure observed in \CCI{} (Fig.~\ref{INS_Tdep}e--f). Another ingredient not included in our model is the pronounced Ising-type anisotropy of the spin-spin correlations in CeCoIn$_5$.
This anisotropy is revealed by polarized neutron scattering measurements of the spin resonance~\cite{raymond2012, raymond2015} and is further supported by the $c$-axis-polarized spin-density-wave character of the ``Q phase''~\cite{kenzelmann2008, kenzelmann2010}. A more detailed modeling and a rigorous treatment of these effects will be necessary to fully capture the details of our observation.

Perhaps the most intriguing question implied by our results is the potential of the FL$^*$ framework to serve as a unifying description of the exotic metallic states that emerge just above the $d$-wave superconducting dome across several material families. Notably, intense, gapped spin-resonance excitations have been reported across heavy-fermion, cuprate, and iron-based superconductors~\cite{sidis2004,lumsden2010,fujita2011,tranquada2014,dai2015,stockert2023}, most exhibiting a remarkably consistent ratio between the resonance energy and the superconducting gap~\cite{yu2009}. This universality may suggest a common organizing principle underpinning these systems and linking the two scales. Within the FL$^*$ framework, such a link arises naturally: both the spin resonance and the superconducting gap originate from a shared hybridized spinon--electron dispersion, developed by the condensation of the $B_1$ chargon field.

The relevance of FL$^*$ is particularly compelling in the case of slightly underdoped cuprates, where it offers a unified treatment of antiferromagnetism, $d$-wave superconductivity, and charge order on the square lattice~\cite{christos2023model}. Notably, it accounts for quantum oscillation measurements~\cite{bonetti2024} and anomalous magnetotransport in the pseudogap regime of cuprates~\cite{fang2022, chan2025, bonetti2025critical}; specifically, the observed Yamaji effect is consistent with the FL$^*$ framework~\cite{YaHuiSS25}. These developments imply that the electronic Fermi surfaces and fractionalized spin excitations of FL$^*$ may underlie a common organizing principle of correlated metals near unconventional superconductivity. In fact, CeCoIn$_5$ closely parallels the cuprates in its layered square-lattice structure, quasi-two-dimensional transport properties, and the presence of a quantum phase transition beneath the superconducting dome~\cite{Petrovic2001,Settai2001,Kohori2001,Sparn2002,Sidorov2002,Bianchi2003,Paglione2003,Christianson2004,Nakajima2007,stock2008,Zhou2013,Tokiwa2013,Allan2013,VanDyke2014,Spehling2009,maksimovic2022}. A key advantage of CeCoIn$_5$ is that this criticality can be accessed in a clean, stoichiometric setting, free from the strong chemical disorder that plagues many cuprates~\cite{Petrovic2001,Bianchi2003}. Nevertheless, extending similar spectroscopic investigations to the cuprates remains highly appealing, especially given their much higher $T_{\mathrm{c}}$ and their central role in the broader context of unconventional superconductivity.


\providecommand{\noopsort}[1]{}\providecommand{\singleletter}[1]{#1}%

\clearpage
\section*{Methods}
\subsection*{Sample preparation}
The inelastic neutron scattering (INS) measurements presented in this study were performed on the same co-aligned CeCoIn$_{5}$ single crystals previously used by Stock et al.~\cite{stock2008}. A total of 5\,g of crystals were co-aligned on multiple thin aluminum plates with the $(HHL)$ plane horizontal. The quality and alignment of the crystal array were verified by a sharp, clean nuclear Bragg peak profile; see Fig.~S2.

\subsection*{Single-crystal inelastic neutron scattering (INS)}
INS data were collected using the CNCS time-of-flight spectrometer at the Spallation Neutron Source, Oak Ridge National Laboratory (ORNL). Measurements were performed with a \textsuperscript{3}He cryostat at temperatures of 0.3, 1.0, 1.6, 2.0, 2.15, 2.5, and 4\,K. The primary measurements were carried out with an incident energy of $E_{\mathrm{i}} = 2.49$\,meV using the high-flux chopper setting operating at $f = 240$\,Hz. The sample was rotated through angular windows ranging from 120$^\circ$ to 360$^\circ$ in total width (which depends on the temperature), with a step size of 1$^\circ$ and a nominal counting time of 4---9 minutes per step (also temperature-dependent). To assess the overall crystal alignment quality, additional measurements were conducted using $E_{\mathrm{i}} = 10$\,meV, which provided nuclear Bragg reflection profiles across a broad region of momentum space (Fig.~S2). The measurements were conducted under the beam power of 1.7\,MW for 0.3\,K, 1.5\,K and 2\,K, and 1.8\,MW for 1\,K, 2.15\,K, 2.5\,K and 4\,K.  

Post-processing and visualization of the INS spectra were carried out using the Mantid~\cite{Mantid} and Shiver~\cite{savici2022_shiver, Shiver} software packages. Background subtraction was performed using data measured at the same temperature in momentum regions free of magnetic signal, as detailed in Supplementary Note~I and Fig.~S3. The datasets were symmetrized using the crystallographic symmetry operations of CeCoIn$_{5}$ to enhance statistical quality. No artifacts are introduced by this treatment; see Supplementary Fig.~S4 for a comparison before and after symmetrization. The integration widths used for the momentum directions perpendicular to the plotted slices are summarized in Table~S1.

\subsection*{Parameters of Model Hamiltonian}

In the minimal model Hamiltonian, Eq.~\eqref{model}, we adopt the tight-binding parameters for the conduction electrons from Ref.~\cite{Allan2013}, which provide a good fit to quasiparticle interference experiments. Retaining hopping processes up to the third nearest neighbors, the parameters are $t_1 \simeq -50.0\,\mathrm{meV}$, $t_2 \simeq -13.34\,\mathrm{meV}$, and $t_3 \simeq 16.7\,\mathrm{meV}$.

The Heisenberg interaction $J_H$ is primarily fixed by the upper energy scale of the DSSF observed in the INS spectrum, which is approximately $2\,\mathrm{meV}$. Within our theoretical framework, this scale corresponds to twice the maximum spinon excitation energy, which in a mean-field description is approximately $1.36\,J_H$. This yields an estimate of $J_H \simeq 0.733\,\mathrm{meV}$. 

The Kondo interaction $J_K$ is constrained by the requirement of a superconducting ground state as well as by the energy of the spin-resonance mode. Based on the detailed calculations presented below, we find that an optimal value of $J_K$ is approximately $60\,\mathrm{meV}$. We note that this magnitude of $J_K$ is comparable to the underlying tight-binding energy scales of the conduction electrons, a condition that is essential for the condensation of the chargon field and the emergence of superconductivity.

\subsection*{Dynamic spin structure factor}

The DSSF contains contributions from both the spin dipole moments of the conduction electrons and the local moments of the $f$ electrons. The spectral weight associated with the $f$ electrons is concentrated at energies of order twice the spinon excitation scale, which lies within the energy window relevant to INS measurements. By contrast, the contribution from conduction electrons is distributed over the much larger kinetic energy scale of the conduction band, yielding a spectral weight that is approximately $1\%$ of that from the $f$ electrons for the model parameters considered here. Consequently, within the energy range of interest, the INS spectra are dominated by the $f$-moment contribution, and we neglect the conduction-electron contribution in the analysis.

Our analysis begins with a mean-field calculation that self-consistently determines the chargon field $B_j$ and the link fields $U_{jj'}$. The interaction between spinons and conduction electrons is mediated by quantum fluctuations of the chargon field, described by its propagator $D(\mathbf{q},\Omega)$. The dynamic spin structure factor is computed using the ladder diagram shown in Fig.~\ref{dssf}b. 

To simplify the calculation, we neglect the explicit momentum and frequency dependence of $D(\mathbf{q},\Omega)$ and replace it by an effective, averaged propagator obtained by integrating over the Brillouin zone and over a frequency window comparable to the spinon excitation scale,
\begin{eqnarray}
D_{\mathrm{eff}} 
= \int \frac{d\mathbf{q}}{(2\pi)^2}
  \int_{-\Lambda}^{\Lambda} \frac{d\Omega}{2\pi}\,
  D(\mathbf{q},\Omega)
= \frac{\alpha}{4}\, J_K ,
\end{eqnarray}
where $\Lambda \simeq 2$meV. This procedure yields $\alpha \simeq 0.58$ at $T=0.13\,\mathrm{K}$. The validity of this approximation is supported by Fig.~\ref{dssf}d, which shows that $D(\mathbf{q},\Omega)$ exhibits only modest variation (approximately $10\%$) over the momentum and frequency range pertinent to the magnetic excitations.

\vspace{1cm}

\section*{Data Availability}
The neutron scattering data presented in this manuscript are publicly available at~\cite{doi_oncat}

\vspace{1cm}

\section*{Code Availability}
Custom codes used in this article are available from the corresponding authors upon request.  

\vspace{1cm}

\begin{acknowledgments}
We thank Filip Ronning for helpful discussions. This work was supported by the U.S. Department of Energy, Office of Science, Basic Energy Sciences, Materials Sciences and Engineering Division. This research used resources at the Spallation Neutron Source, a DOE Office of Science User Facility operated by the Oak Ridge National Laboratory. The beam time was allocated to CNCS on proposal number IPTS-32607.1 and IPTS-35088.1. 
S.S. and P.M.B. were supported by NSF Grant DMR-2245246 and by the Simons Collaboration on Ultra-Quantum Matter which is a grant from the Simons Foundation (651440, S. S.). 
The Flatiron Institute is a division of the Simons Foundation.
P.M.B. acknowledges support by the German National Academy of Sciences Leopoldina through Grant No.~LPDS 2023-06 and the Gordon and Betty Moore Foundation’s EPiQS Initiative Grant GBMF8683. Part of this work was carried out at the Brookhaven National Laboratory which is operated for the U.S. Department of Energy by Brookhaven Science Associates (DE-AcO2-98CH10886).
\end{acknowledgments}

\section*{Author contributions}
P.P., S.S., C.D.B., and A.D.C. conceived the project. C.P. grew the single crystals. C.S. prepared the co-aligned sample array for neutron scattering. P.P., A.A.P., D.M.P., M.B.S., and A.D.C. conducted the neutron scattering measurements. P.P. analyzed the neutron scattering data. S.-S.Z., P.M.B., S.S., and C.D.B. conceived and performed the theoretical analysis. P.P., S.-S.Z., C.D.B., and A.D.C. wrote the paper with contributions from all authors.

\section*{Competing Interests}
The authors declare no competing interests. 

\vspace{1cm}



\clearpage
\end{document}